# On the importance of light scattering for high performances nanostructured antireflective surfaces

*Florian Maudet[1†*], Bertrand Lacroix[2,3], Antonio J. Santos[2,3], Fabien Paumier[4*], Maxime Paraillous[5], Simon Hurand[4], Alan Corvisier[4], Cyril Dupeyrat[5], Rafael García[2,3], Francisco M. Morales[2,3] and Thierry Girardeau[4]*

1. Institute Functional thin film oxides for energy-efficient future information technology, Helmholtz-Zentrum Berlin für Materialien und Energie Hahn-Meitner-Platz 1, 14109, Berlin, Germany
2. Department of Materials Science and Metallurgic Engineering, and Inorganic Chemistry, Faculty of Sciences, University of Cádiz, Spain.
3. IMEYMAT: Institute of Research on Electron Microscopy and Materials of the University of Cádiz, Spain.
4. Institut Pprime, UPR 3346 CNRS-Université de Poitiers-ENSMA, SP2MI, 86962 Futuroscope-Chasseneuil cedex, France
5. Safran Electronics and Defense, 26 avenue des Hauts de la Chaume, 86280 Saint-Benoît, France

An antireflective coating presenting a continuous refractive index gradient is theoretically better than its discrete counterpart because it can give rise to a perfect broadband transparency. This kind of surface treatment can be obtained with nanostructures like moth-eye. Despite the light scattering behavior must be accounted as it can lead to a significant transmittance drop, no methods are actually available to anticipate scattering losses in such nanostructured antireflective coatings. To overcome this current limitation, we present here an original way to simulate the scattering losses in these systems and routes to optimize the transparency. This method, which was validated by a comparative study of coatings presenting refractive indices with either discrete or continuous gradient, shows that a discrete antireflective coating bilayer made by oblique angle deposition allows reaching ultra-high mean transmittance of 98.97% over the broadband [400-1800] nm. Such simple surface treatment outperforms moth-eye architectures thanks to both interference effect and small dimensions nanostructures that minimize the scattering losses as confirmed by finite-difference time domain simulations performed on reconstructed volumes obtained from electron tomography experiments.

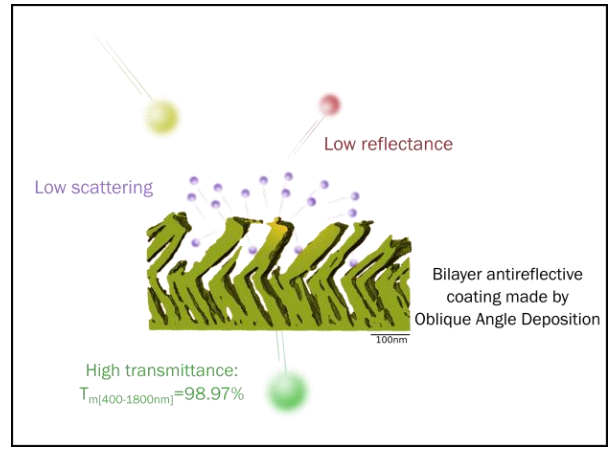

## I. Introduction

Antireflective coatings (ARC) are indispensable features of high transmittance substrate with applications in a wide variety of domains of photonics as for example in optics or photovoltaic.[1–4] Particularly the need for a broadband ARC ensuring a high transparency glass substrate from the visible up the near infrared (NIR) is of interest accordingly to the available detectors.[5] Different approaches can be used to achieve broadband ARC. So far, the main way relies on the deposition of thin films forming a multilayer stack alternating materials with high and low refractive indices.[6,7] The refractive index (RI) and thickness of each layer are optimized to create destructive interference with the reflected wave.[6,7] However, for broadband wavelength applications this approach is challenging due to the lack of material available and to the complexity of the layer assembly to deposit in order to obtain high transmittance.[3,6,7] To go beyond these limitations, other approaches have been considered. As it is now well established, a gradient refractive index ARC (GRIARC) can suppress efficiently the light reflection over wide spectral ranges.[8,9] Recently, K-H. Kim and Q-H. Park have demonstrated that it is indeed possible to perfectly remove the light reflection of any wavelength using a gradient profile of refractive index determined analytically.[10] From this analytical development it was established that perfect broadband antireflective property comes with a constraint on the dispersion law of the coating material. For thin GRIARC nanostructures in regard to the probing wavelength, it can be shown that perfect reflectance suppression comes with the use of anomalous dispersion materials. In order to obtain perfectly transparent ARC, this has to be avoided since the anomalous dispersion behavior of a material usually comes from an increasing extinction coefficient that will consequently lead to light absorption.[10] It may be noted however, that the use of new metamaterials, still in the early development stage, could remove this constraint.[11]

Nevertheless, the dispersion law becomes normal for sufficiently thick GRIARC structures. The optimal gradient profile is then close to the well-known quintic profile that was determined from numerical optimizations.[12] It can be derived from those calculations that using normal dispersion law material there will be a thickness dependency of the ARC efficiency.

In practice these ARC can be achieved by employing innovative nanostructuration methods to reduce the refractive index through the introduction of porosity, as conventional dense materials with RI below 1.39 do not exist.[3] GRIARC can be made by moth eye nanostructures (ME), called from their natural presence on the eyes of moths.[13] These biomorphic structures, consisting in pseudo-regular arrays of nanopillars, are mainly characterized by two parameters: their height, that is equivalent to the GRIARC thickness, and their width. Due to their shape, these ME structures present a gradually increasing porosity from their bases to their tips. If their dimensions are sufficiently small in regard of the probing light wavelength, the mixing result can be considered as a gradient effective RI that can be calculated from the Bruggeman effective medium approximation (BEMA).[14] Such structures are usually obtained by a reactive ion etching process although it can also be made from embossing process or other replicative methods. [13,15,24,16–23] These GRIARC follow behavior predicted by the BEMA for the long wavelength range, a thicker GRIARC is more efficient to suppress reflectance than its thinner counterpart[1,15,17,19,20,22,25]. On the contrary, on the short wavelength range a thicker GRIARC leads to higher optical losses that reduce the transmittance level[22]. This results in a necessary compromise to be found on the optimal thickness. Such compromise is usually determined empirically as the optical losses are not considered by the BEMA model.

Furthermore, it was pointed out that when a very low RI, or a high level of porosity, cannot be obtained, the performances of a discrete ARC can be higher than a continuous gradient thanks to the interference effects.[26] Such discrete ARC, can be made using oblique angle deposition (OAD) as it is a viable, flexible technique among all existing fabrication processes.[3,8,13,26–30] It allows a fine-tuning of the porosity inserted at the nanoscale through the adjustment of the deposition angle α, thus promoting a high control of the RI.[31]

In this article we aim to compare the transmittance performances of GRIARC with a discrete ARC made by OAD on the broadband range of [400-1800] nm. We specifically focus on the light scattering that may arise in these nanostructures as it hinders the transparency of thick ARC. An original method is presented to calculate those optical losses, making easier future designs and optimizations of nanostructured discrete ARC or GRIARC. Based on this approach, a discussion is made to understand the current limits of broadband nanostructured ARC performances and what problems need to be solved to lift them.

## II. Experimental Section

The experimental setup used for the oblique angle deposition consists in a vacuum chamber (base pressure $2.10^{-6}$ mbar) equipped with an electron-beam evaporator. For the calibration step of thickness and refractive index, $SiO_2$ OAD layers were deposited at various angles of incidence from 0° to 85° by e-beam evaporation method using a $SiO_2$ crucible (Umicore© purity>99.99%) on 1 mm thick, 1 inch diameter BK7 glass substrates. A quartz crystal monitor was used to control the thickness and assure a deposition rate of 10 Å/s in normal incidence. The deposition was performed at room temperature. This elaboration set-up is not equipped with an in-situ adjustable orientation sample-holder, so the bilayer was made in two steps with an azimuthal rotation of Φ=180° in between. For the first step, a $SiO_2$ layer was deposited at an angle of 65° with the condition of elaboration previously described. For the second step, the chamber was opened to modify the angle of incidence to a value of 85°. The second layer of $SiO_2$ was then evaporated and deposited onto the first OAD layer. This whole process was repeated a second time to coat the BK7 substrate on both sides. Note that this bilayer was deposited at the same time on a (001) silicon substrate, which was intentionally placed with one <110> crystallographic axis perpendicular to the incoming deposition flux for further analyses by transmission electron microscopy (TEM) [32].

To access the optical properties of monolayer and bilayer, spectrophotometry measurements were carried out for each sample. Transmittance and reflectance spectra were recorded with a Carry 5000 Varian spectrophotometer. Transmittance and reflectance acquisitions were repeated 4 times for each sample giving a repeatability of 0.1%. As anisotropic behavior may be expected from OAD nanostructures, measurements were made with a polarization in the direction of the columns. The nanostructure of the OAD samples was analyzed by TEM experiments at an acceleration voltage of 200 kV. Bright-field TEM images were recorded in a JEOL 2010F microscope. The 3D morphology of the OAD $SiO_2$ bilayer was extracted from electron tomography in a FEI Titan Cubed Themis 60-300 microscope, using the high-angle annular dark-field scanning TEM (HAADF-STEM) imaging mode. For that purpose, a dedicated tomography holder operated by the FEI Xplore3D software was used to acquire tilt series every 2° from -60° to +60°. The FEI Inspect 3D software was then employed to align the projections using cross-correlation methods, and to perform reconstruction into a 3D volume using the conventional simultaneous iterative reconstruction technique (SIRT) with 30 iterations. The FEI Avizo program was used for 3D visualization and segmentation of the reconstructed volume. Specimen for TEM were prepared by soft and flat mechanical polishing of the two faces using a tripod apparatus (Model 590 Tripod Polisher) in order to control finely the thinning down to a few microns. This step was followed by a short $Ar^+$-ion milling step in a Gatan PIPS system (using an energy of 3.5 keV and an incidence angle of +/-7º). This procedure provides large and homogeneous electron transparent areas, which is suitable to limit the geometrical shadowing of the specimen at large tilt angles of the holder. Prior to insertion into the microscope, the specimen was cleaned in an $Ar/O_2$ plasma in order to remove the hydrocarbon contamination.

Finite-difference time domain (FDTD) simulations were carried out with the Lumerical© commercial software. The 3D reconstruction model of the $SiO_2$ bilayer obtained from electron tomography was slightly cropped to remove defects on the edges of the reconstruction and then converted into a surface file to be imported in the FDTD software. We then assumed that the refractive indexes of the material and porosity (empty space) components in the 3D reconstructed volume correspond to that of the dense $SiO_2$ film and of air, respectively. A simulation box of 800x50x500 $nm^3$ with perfect matched layer boundaries and a mesh size of 1 nm was used. Total Field/Scattered Field source with slightly higher dimensions than the 3D bilayer was used to determine the scattering cross section of the structure. The source polarization was aligned with the direction of the columns to respect the spectrophotometric measurements conditions.

The ME simulations were performed under the same conditions with a box size varying accordingly to the dimensions of the pyramid structures. Five height dimensions of pyramid going from 500 to 2000 nm were tested with each 8 square bases dimensions ranging from 10 to 200 nm.

### III. Results and discussions
#### A. Different theoretical approaches for broadband antireflective coatings

As previously stated, a GRIARC needs to be sufficiently thick to be efficient over a broadband. To emphasize this aspect, we present on Figure 1.a the simulated transmittance spectra on the wide [400-2000] nm range for different thicknesses of GRIARC on both sides of a glass (BK7) substrate following the quintic profile. The optical designs of the GRIARC systems are given on Figure 1.b.

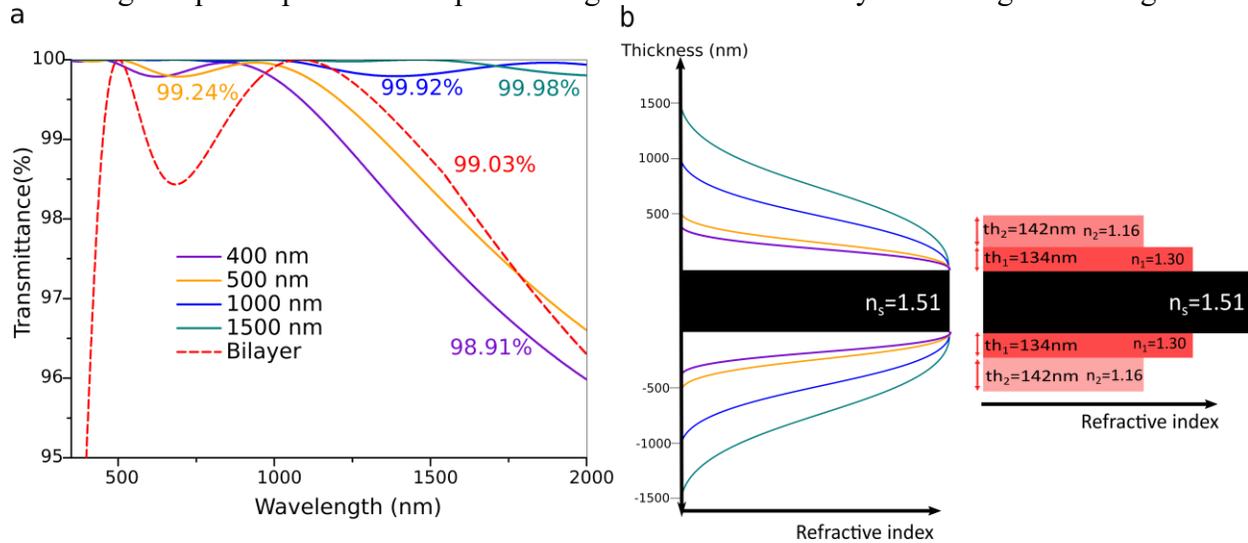

Figure 1 a. Transmittance spectra of GRIARC coatings of different thicknesses with a quintic profile compared with a bilayer interferential coating. The mean transmittance value of each spectrum is presented for the [400-1800] nm range. b. Schematic representation of the optical design of the different coatings: GRIARC designs on the left, bilayer design on the right.

Although thin GRIARC (400 and 500 nm) are relatively efficient for short wavelengths, it can be observed that the transmittance in the NIR wavelength range is limited by reflection of the light. To minimize optical losses in this domain, thicker GRIARC must be considered as demonstrated from simulations of the thicknesses of 1000 and 1500 nm. Actually, to obtain at least 99% mean transmittance over the 400-1800 nm range ($T_{m[400-1800] nm}$) a GRIARC coating thicker than 400 nm must be used.

The simulation of an optical design consisting in an interference-based bilayer implementing fictive values of refractive index was also performed and compared to the GRIARC systems. Those values are considered as fictive as they are free parameters of the simulation and might not correspond to the refractive indices of existing bulk materials. This numerically optimized bilayer is thinner (total thickness of 276 nm) than the presented GRIARC systems. It also presents high transmittance ($T_{m[400-1800 nm]}$=99.03%), notably because it benefits from the interferential effect in the NIR range.

### B. Experimental comparison of the transmittance performances of GRIARC and discrete ARC coatings

Because this bilayer design presents the appeal of simplicity and a high transmittance level, it has then been prepared using oblique angle deposition (OAD) technique. Prior to the fabrication of the bilayer, a preliminary step was devoted to the calibration of the refractive index and thickness of individual $SiO_2$ layers prepared by OAD at different deposition angles. After finding out the optimal deposition parameters, the ARC bilayer was then deposited on a BK7 substrate in a two-step process. First, a $SiO_2$ layer was deposited at α=65° with a targeted thickness of 134 nm and refractive index of n=1.30. Then a second layer of $SiO_2$ was deposited at α=85° with a targeted thickness of 142 nm and refractive index of n=1.16. An azimuthal rotation of 180° was performed in between the layers in order to minimize any material deposition within the porosity of the first layer during the deposition of the second one. A TEM bright-field micrograph of the resulting stack is presented on Figure 2.a. It can be observed that the obtained thicknesses are close from the targeted ones respectively 143 nm and 161 nm for the first and the second layer. As expected, the nanocolumns in the top layer are more tilted, yielding to higher porosity due to geometrical shadowing. This micrograph also clearly indicates that the OAD nanocolumns of the first layer act as nucleation sites for the second one. According to Figure 2.b and Figure 2.c, this simple bilayer ARC coating presents outstanding performances ($T_{m[400-1800] nm}$=98.97%). The resulting transmittance and reflectance spectra match almost perfectly with the design counterparts. Interestingly, the results are quite close to the performances obtained by Sood *et al.* demonstrating the reproducibility of this method.[2] The small mismatch present in the short wavelength range (below 500 nm) between the experimental and designed ARC coatings will be discuss further after. To go deeper, we have compared the performance of our OAD bilayer with that of experimental GRIARC system based on ME structures according to Z.Diao *et al.* ( Figure 2.b and Figure 2.c).[22]

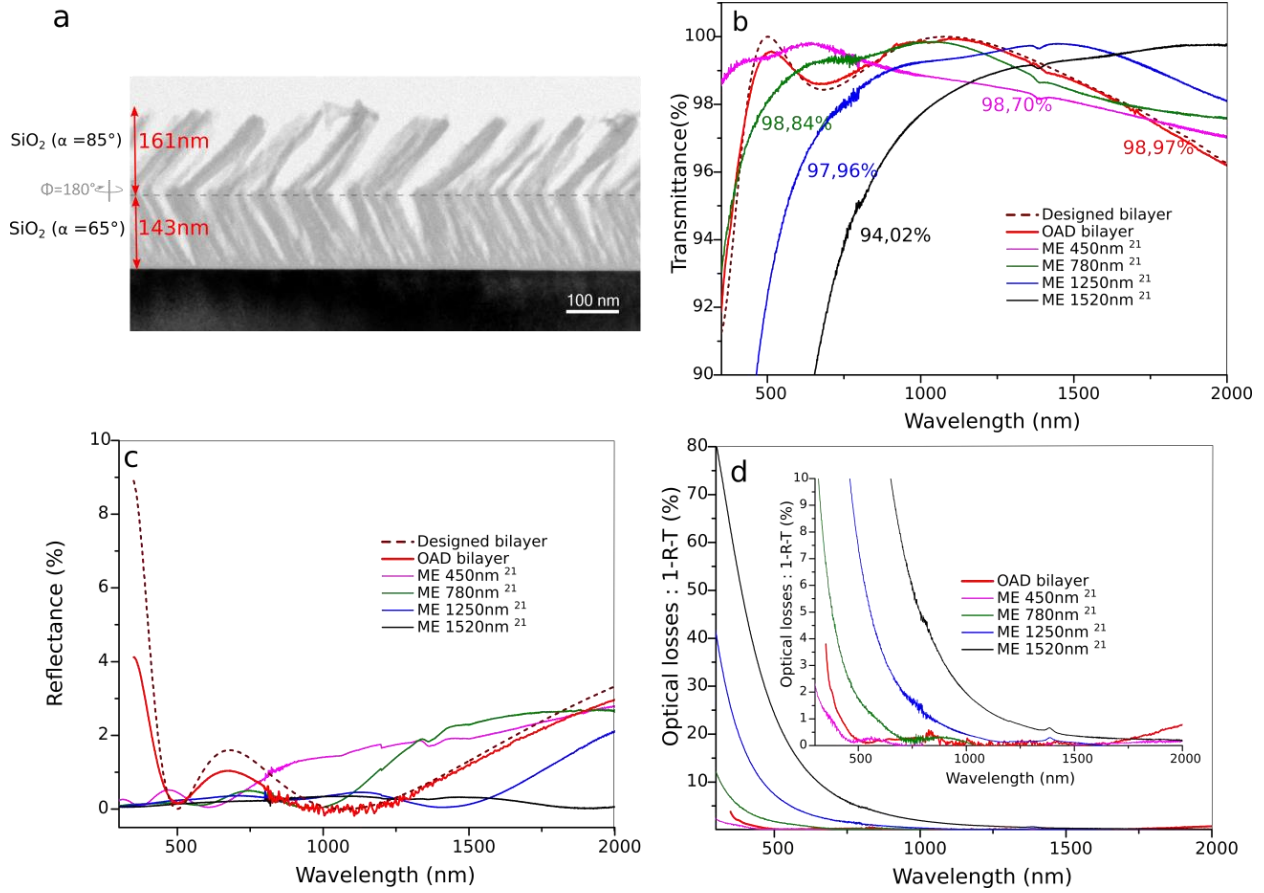

Figure 2 a. TEM bright field micrography of the ARC OAD bilayer composed of $SiO_2$ stacks deposited at α=65° and α=85° with azimuthal rotation between the two layers of Φ=180°. The observation was done along a <110> zone axis of the Si susbtrate. Transmittance (b), Reflectance (c) and optical losses (d) spectra of the OAD bilayer deposited on a glass substrate BK7 on two sides. The dashed lines in figure b and c present the transmittance and reflectance spectra of the optical design of the bilayer. The transmittance, reflectance and optical losses spectra of ME measured by Z.Diao et al.[22] were added on respectively figure b, c and d. The mean transmittance value of each spectra is presented for the [400-1800] nm range.

As predicted by the theory and demonstrated on Figure 1, the use of thicker ME structures minimizes the reflectance thus increasing the transmittance in the NIR range as observed on Figure 2.c. However in counterpart, the transmittance level is lowered in the visible range for an increasing thickness of the ME (Figure 2.b), a behavior not described by BEMA simulations (Figure 1.a). This compromise between short and long wavelengths allows reaching a mean transmittance $T_{m[400-1800]\,nm}$=98.84% for the best GRIARC nanostructures, below that of the OAD bilayer prepared in this work.

To study the origin of low light transmittance in the short wavelength range, optical losses of the different systems (GRIARC and OAD bilayer) are presented on Figure 2.d. The optical losses considered here are deduced from transmittance and reflectance measurements as follows:

$$Optical\ losses(\lambda) = 1 - R(\lambda) - T(\lambda) \quad (1)$$

where R and T denotes respectively the reflectance and transmittance.

The OAD bilayer as well as the 450 nm ME present a low level of optical losses mostly in the UV range. However, thicker ME structures exhibit an increasingly high level of optical losses in this spectral range up to 80% at λ=350 nm for a thickness of 1520 nm. Therefore, optical losses are responsible for the weak transmittance in the short wavelength range for thick ME GRIARC. This demonstrates that the performances of ARC nanostructures cannot solely be determined from reflectance measurements and that optical losses by absorption or scattering must be assiduously considered.

As pointed out by Niklasson *et al.*, a scattering behavior can be observed when the dimensions of the nanostructures are non-negligible in regards of the probing wavelength according to the Mie theory.[14] This therefore limits the validity of the BEMA for objects of small sizes and may explain the difference observed between designs and experiments in the short wavelength range. This point is supported by the diffuse scattering reported in the IR range for thick (several microns) ME structures.[33]

### C. A method to simulate scattering losses from nanostructured ARC

In order to consider the scattering behavior of the ARC nanostructures and anticipate the optical losses observed in the UV-Vis range, we propose a method based on finite-difference time domain simulations and the Mie theory. According to Mie theory for a transparent medium, in absence of multiple scatterings (which is the case of weak scattering structures), the scattering efficiency $Q_{scat}$ is related to the scattered intensity by: [34]

$$I_{scat}(\lambda) = 1 - I_{trans}(\lambda) = 1 - te^{-Q_{scat}(\lambda)} \quad \text{with} \quad Q_{scat}(\lambda) = \frac{\sigma_{scat}(\lambda)}{S} \quad (2)$$

Where $\sigma_{scat}$ is the scattering cross section and S is the section area. The scattering cross section can be determined from FDTD simulations using total-field/scattered-field (TFSF).[35] Note that for the sake of simplicity, since the structures have a very low reflectivity, we only consider here the light attenuation due to scattering for the transmitted wave. In the case of a substrate coated on both sides, if we neglect light reflection at the interfaces, the corresponding total scattered intensity can be calculated as:

$$I_{scat}(\lambda) = 1 - e^{-2Q_{scat}(\lambda)} \quad (3)$$

### D. Simulated and experimental scattering of nanostructured ARC

Firstly, we applied this method to estimate the on the optical losses of the ARC OAD bilayer. To simulate accurately the scattering cross section by FDTD, the complex 3D geometry of this system was determined experimentally from electron tomography (see Figure 3.a). The reconstructed volume, that includes morphological information of the coating of a wide spatial range (mostly the distribution of material and porosity components), was then used as an input parameter for the calculations, considering that the columns have the dielectric permittivity of $SiO_2$. This simulated scattered intensity is presented on Figure 3.a along with the experimental optical losses measured for this ARC OAD bilayer.

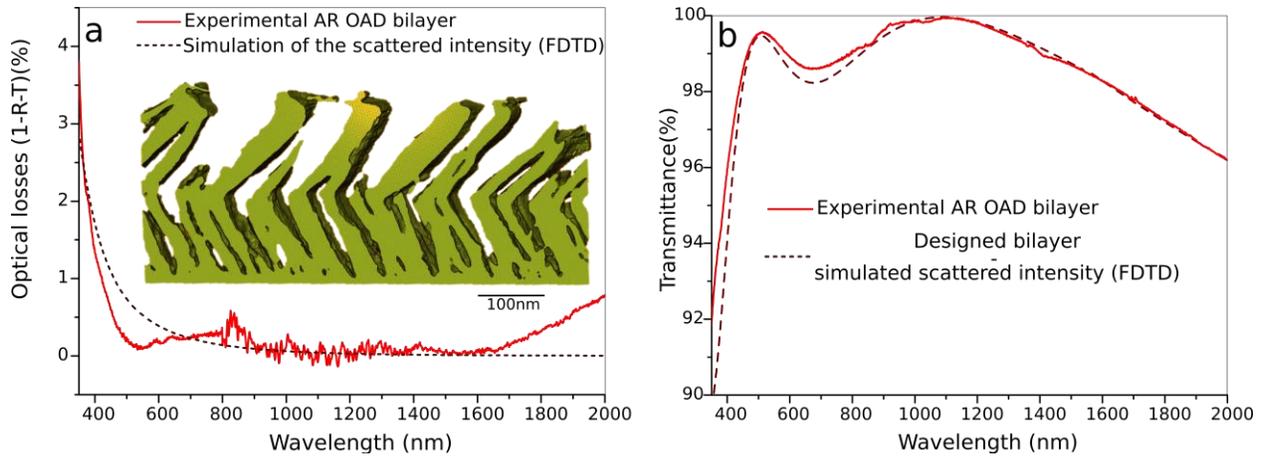

Figure 3 a. Optical losses of the $SiO_2$ OAD bilayer deposited on glass substrate measured experimentally (full line) and calculated from FDTD using the 3D reconstruction obtained from electron tomography (dashed line). b. Transmittance spectra of the OAD bilayer measured experimentally (full line) and calculated by subtracting the simulated scattered intensity presented on figure a to the designed value (dashed line).

The very good agreement between the experimental and simulated spectra tends to confirm that this innovative approach combining the implementation of realistic 3D sample geometry with advanced simulation of light propagation is appropriate to account for optical losses. This result also validates that the optical losses can be attributed to the optical scattering of the nanostructures. Knowing the scattering losses, we present on the Figure 3.b the design spectra minimized from the simulated scattered intensity along with the experimental spectra of the OAD bilayer. It is possible to attribute the small mismatch between the designed spectra and the experimental OAD bilayer (pointed out before on Figure 2.b) to the scattering behavior of the ARC. Thus, the OAD bilayer forms a low scattering nanostructured film thanks to the small radius of the columns and its small thickness.

For comparison, we used the same method to evaluate the scattering behavior of $SiO_2$ ME nanostructures. For the sake of simplicity, these structures were approximated by square pyramids of width w and thickness t (as described on Figure 4.a). The scattered intensity was then deduced using equation (3) from the Mie scattering cross section for different thicknesses and a constant aspect ratio of $r = \frac{t}{w} = 10$. The resulting spectra are presented on Figure 4.a.

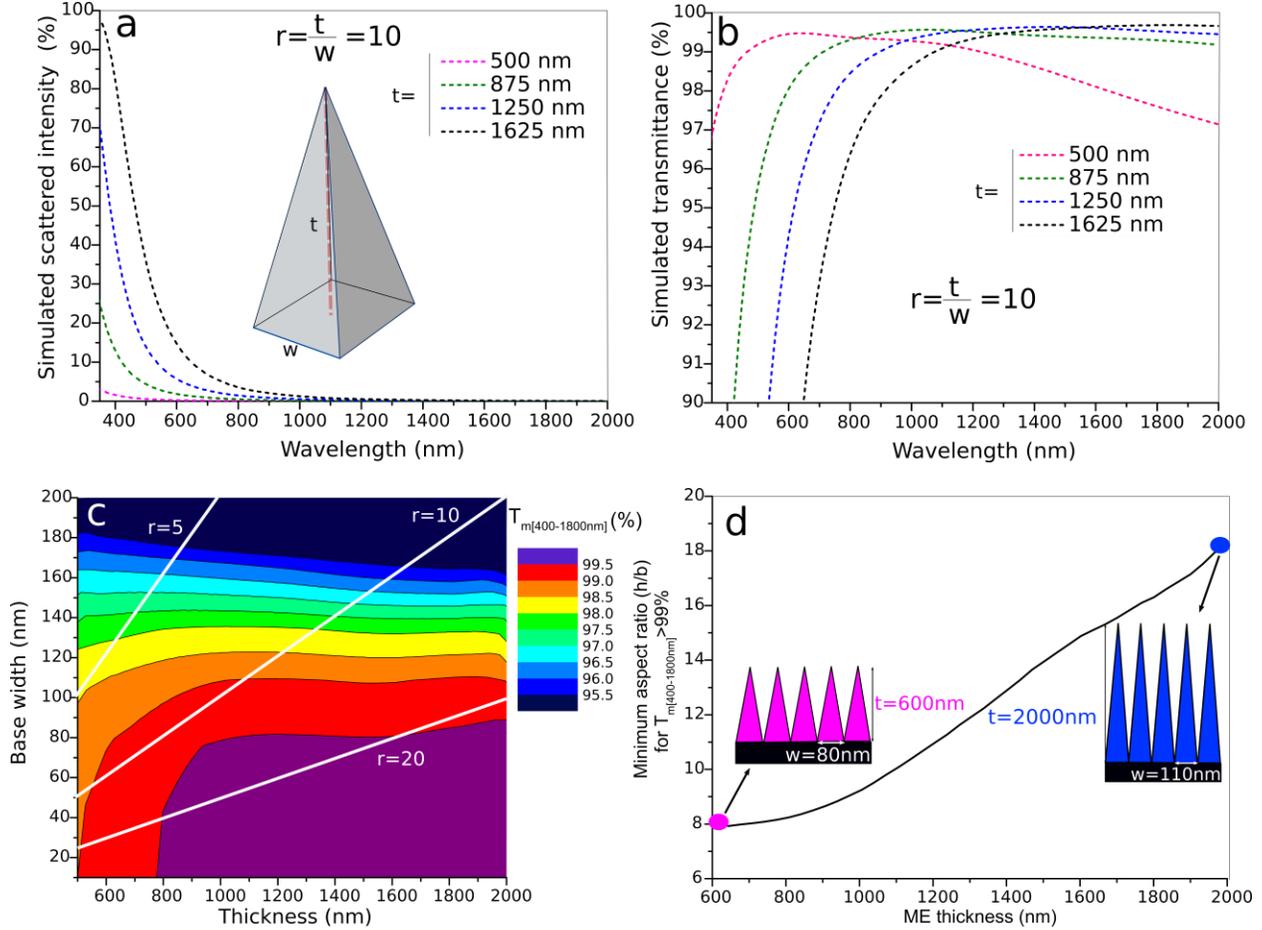

Figure 4 a. Simulated scattered intensity spectra from FDTD Mie scattering cross section of $SiO_2$ ME pyramid nanostructures with an aspect ratios of r=10 for different thicknesses b. Simulated transmittance of spectra $SiO_2$ ME pyramid nanostructures simulated by the BEMA for different thicknesses subtracting the scattered intensity presented in Figure 4.a c. Contour plot of $T_{m[400-1800nm]}$ simulated from the BEMA and taking into account the scattered intensity for pyramid ME of different thicknesses t and base widths d. Minimum aspect ratio for $T_{m[400-1800]\,nm}$=99% for different thicknesses of ME $SiO_2$ pyramid nanostructures.

This figure shows that, for a constant aspect ratio of 10 and an increasing thickness of ME structures, the scattered intensity dramatically increases from a maximum value of 1.5% for a thickness of 500 nm up to 96% for 1625 nm. Interestingly the simulated scattered intensity reveals a similar behavior than the optical losses of ME nanostructures presented in Figure 2.d. This also confirms the scattering nature of the optical losses.

If we consider a GRIARC constituted of perfect periodic square based pyramids, the porosity for any position z, the vertical axis, can be calculated as:

$$P(z) = \left(1 - \frac{z}{t}\right)^2 \quad (4)$$

where t is the total thickness of the structure.

BEMA can be applied from this porosity profile by decomposing it to 100 layers of increasing porosity rate in order to calculate the transmittance spectra of ME square based pyramids. Then by combining the scattering results to BEMA simulations, we calculated the transmittance of ME structures. For that, both transmittance (BEMA) and scattering properties (FDTD) for different thicknesses of ME for a constant aspect ratio of r=10 were considered.

The similarity between these results (Figure 4.b) and experimental transmission (Figure 2.b) confirms the validity of the approach adopted in this paper. This optical behavior, that consists in a transmittance diminution in the UV range coupled with an increasing transmittance in the IR range for an increasing thickness, is consistent with what can be found in literature.[1,15,17,19,20,22,25]

To better understand the link between ME dimensions and the scattering losses, a contour plot of the mean transmittance $T_{m[400-1800]\,nm}$ of ME with different thicknesses and widths is presented on Figure 4.c. This figure is presented for a minimum thickness value of 500 nm because below, the GRIARC is not efficient to suppress long wavelength reflection in order to obtain a minimum of 99% mean transmittance. To facilitate the discussion three straight lines have been added corresponding to three different values of the constant aspect ratio: r=5, 10 and 20. So Figure 4.b corresponds to a variation of thickness on the straight lines r=10.

Figure 4.c shows that the width of the ME is the critical parameter to adjust in order to obtain a high broadband transmittance. A maximum width of 110 nm is required to obtain $T_{m[400-1800\,nm]}$ >99%. If this width is kept constant, the performances are barely affected by an increase of the thickness if the GRIARC is sufficiently thick (thicker than 900 nm). So an increasing thickness will be beneficial only if the width is kept constant *i.e.* only if the aspect ratio increases. In other words, for an increasing thickness, an increasing aspect ratio of ME is needed to maintain a high $T_{m[400-1800\,nm]}$.

### E. Discussion

Since the $T_{m[400-1800\,nm]}$ is equal to 99% for the OAD bilayer, we present on Figure 4.d the minimum aspect ratio of needed to obtain this mean transmittance for different thicknesses of pyramids ME structures. Interestingly, pyramid ME nanostructures with a thickness of 600 nm and an aspect ratio of 8 will have the same mean transmittance (99%) than pyramid ME of 2000 nm thickness and an aspect ratio of 18. Consequently, enhancing the mean transmittance of ME systems above 99% is in principle possible to the constraint of nanostructuring very high aspect ratio and thick architectures. However, beyond the mechanical instability that might be associated with high aspect ratio ME, such nanostructuration is challenging from the technological point of view. Indeed, the etching process usually employed to realize ME offers a limited control on the width of the structures. To increase the ME aspect ratio, it is necessary to minimize the dimension of the shadowing elements although these are already quite small. For example, Z. Diao *et al* reported the use of gold nanoparticles of 6 nm diameter as masks.[22] This therefore demonstrates clearly the interest of the OAD approach in the manufacturing of highly efficient antireflective coatings. By taking advantage of the interferences, the ARC bilayer coating developed in this work permits an efficient suppression of the reflectance with relatively small dimensions nanostructures thus avoiding undesirable light scattering. One could think of enhancing the performances of the OAD bilayer by adding supplementary layers. However, it is high likely that such stack will suffer from the scattering losses as the nanocolumns obtained from OAD deposition tend to broaden with the thickness.[36,37] The amelioration of optical ARC properties by OAD is feasible in a near future but will probably require the use of modified growth procedure to minimize the columns dimensions.[38]

## IV. Conclusions

In summary, this paper demonstrates that light scattering is an important aspect to consider for the development of high transmittance ARC surfaces. A simple method to calculate light scattering losses of nanostructured coatings have been introduced. A detailed comparison of the performances of GRIARC and discrete gradient ARC was done to evaluate those optical losses. We showed that a coating with a discrete gradient refractive index presents a very high transmittance level for a relatively small thickness by taking advantage of the interferences. This bilayer prepared by OAD deposition exhibits a mean transmittance $T_{m[400-1800\ nm]}$= 98.97% very close to the designed value. A very small amount of optical losses was measured (around 2% at 400 nm) and attributed to light scattering. The scattering behavior of this bilayer was confirmed from FDTD simulation which was performed on a 3D volume reconstruction obtained from electron tomography. On the other hand, continuous GRIARC nanostructures can in theory result in almost perfect transmittance. However, to be efficient on the long wavelength range, it is shown that these systems must be sufficiently thick, at least 500 nm, to reach mean transmittance as high as 99%. In practice, thick ME nanostructures will favor scattering losses mainly due to the too large width of those structures. In order to limit this effect and keep on enhancing the transmittance, thick GRIARC nanostructures with a high aspect ratio must be manufactured. For technological reasons, this constraint represents a real challenge that may be hard to overcome. Therefore, considering the actual state-of-the-art of nanostructuring methods, development based on discrete gradient ARC should be preferred in the future as they are simpler to produce and provide better optical performances.

AUTHOR INFORMATION

**Corresponding Author**

* E-mail: florian.maudet@helmholtz-berlin.de

* E-mail: fabien.paumier@univ-poitiers.fr

**Present Addresses**

†Helmholtz-Zentrum Berlin für Materialien und Energie
Hahn-Meitner-Platz 1
14109 Berlin
Germany

**Author Contributions**

The manuscript was written through contributions of all authors. All authors have given approval to the final version of the manuscript.

ACKNOWLEDGMENT

This work was supported by the DGA (Direction Générale de l'Armement), the French Defense Procurement Agency. This work has been partially supported by « Nouvelle Aquitaine » Region and by European Structural and Investment Funds (ERDF reference P-2016-BAFE-209): IMATOP project. The "Talent Attraction Program" of the University of Cádiz is acknowledged for supporting B. Lacroix contract code E-11-2017-0117214. A. J. Santos also thanks the financial support of the IMEYMAT Institute as well as the Ministerio de Ciencia, Innovación y Universidades and Ministerio de Educación y Formación Profesional in Spain for the concessions of grants (ICARO-173873 and FPU16-04386)